\documentclass[aps,pra,a4paper,floatfix,twocolumn,superscriptaddress]{revtex4}
\pdfoutput=1
\usepackage{epsfig,amsbsy,bm}
\usepackage{etoolbox}
\usepackage{color}
\usepackage{placeins}
\usepackage{mathrsfs}
\usepackage[colorlinks,linkcolor=red,citecolor=blue]{hyperref}
\usepackage{amsmath}
\usepackage{lipsum}
\usepackage{graphicx}
\usepackage{dcolumn}
\usepackage[nice]{nicefrac}
\usepackage[tight]{units}
\usepackage{lipsum}
\usepackage[dvipsnames]{xcolor}
\usepackage{pgfplots}
\usepackage[utf8]{inputenc}
\usetikzlibrary{patterns}
\usetikzlibrary{arrows.meta}
\usetikzlibrary{fadings}
\usetikzlibrary{spy}
\tikzset{>={Latex[width=2mm,length=2mm]}}
\pgfplotsset{compat=newest}
\pgfplotsset{plot coordinates/math parser=false}
\def\t0{\mbox{$t_{\mbox{{\tiny {0}}}}$}}

\def\p0{\mbox{$p_{\mbox{{\tiny {0}}}}$}}

\def\E0{\mbox{$E_{\mbox{{\tiny {0}}}}$}}

\newcommand{\be}{\begin{equation}}
\newcommand{\ee}{\end{equation}}
\newcommand{\ba}{\begin{eqnarray}}
\newcommand{\ea}{\end{eqnarray}}

\newcommand{\lund}{Department of Physics, Lund University, P.O. Box 118, SE-22 100 Lund, Sweden}

\newcommand{\sthlm}{Department of Physics, Stockholm University, SE-106 91 Stockholm, Sweden}
\newcommand{\gu}{Department of Physics, Gothenburg University, Origov\"agen 6B, SE-41 296 G\"oteborg, Sweden}

\begin{document}
\bibliographystyle{unsrt}
\title{Photoionization in the time and frequency domain}

\author{M. Isinger}\affiliation{\lund}
\author{R.J. Squibb}\affiliation{\gu}
\author{D. Busto}\affiliation{\lund}
\author{S. Zhong}\affiliation{\lund}
\author{A. Harth}\affiliation{\lund}
\author{D. Kroon}\affiliation{\lund}
\author{S. Nandi}\affiliation{\lund}
\author{C. L. Arnold}\affiliation{\lund}
\author{M. Miranda}\affiliation{\lund}
\author{J.M. Dahlstr\"om}\affiliation{\lund} \affiliation{\sthlm}
\author{E. Lindroth}\affiliation{\sthlm}
\author{R. Feifel}\affiliation{\gu}
\author{M. Gisselbrecht}\affiliation{\lund}
\author{A. L'Huillier}\affiliation{\lund}

\date{\today}

\begin{abstract}
Ultrafast processes in matter, such as the electron emission following light absorption, can now be studied using ultrashort light pulses of attosecond duration ($10^{-18}$s) in the extreme ultraviolet spectral range. The lack of spectral resolution due to the use of short light pulses may raise serious issues in the interpretation of the experimental results and the comparison with detailed theoretical calculations.
Here, we determine photoionization time delays in neon atoms over a 40 eV energy range with an interferometric technique combining high temporal and spectral resolution. We spectrally disentangle direct ionization from ionization with shake up, where a second electron is left in an excited state, thus obtaining excellent agreement with theoretical calculations and thereby solving a puzzle raised by seven-year-old measurements. Our experimental approach does not have conceptual limits, allowing us to foresee, with the help of upcoming laser technology, ultra-high resolution time-frequency studies from the visible to the x-ray range.
\end{abstract}

\maketitle

%\section{Introduction}
While femtosecond lasers allow the study and control of the motion of nuclei in molecules, attosecond light pulses give access to even faster dynamics, such as electron motion induced by light-matter interactions \cite{KrauszRevModPhys2009}. During the last ten years, seminal experiments with sub-femtosecond temporal resolution have allowed the observation of the electron valence motion \cite{GoulielmakisNature2010}, monitoring of the birth of an autoionizing resonance \cite{GrusonScience2016,KaldunScience2016} and tracking the motion of a two-electron wavepacket \cite{OttNature2014}, to cite only a few examples. Fast electron motion occurs even when electrons are directly emitted from materials upon absorption of sufficiently energetic radiation (the photoelectric effect). The time for the photoelectron emission \cite{CavalieriNature2007}, which, in free atoms, represents the time for the electron to escape the potential, called photoionization time delay \cite{SchultzeScience2010,KlunderPRL2011}, is typically of the order of tens of attoseconds, depending on the excitation energy and on the underlying core structure.

Photoemission has traditionally been studied in the frequency domain, using high-resolution photoelectron spectroscopy with x-ray synchrotron radiation sources, and such methods have provided a detailed understanding of the electronic structure of matter \cite{SchmidtRPP1992,Becker2012}. Absorption of light in the 60-100 eV range by Ne atoms, for example, leads to direct ionization in the $2\mathrm{s}$ or $2\mathrm{p}$ shells and to processes mediated by electron-electron interaction, leaving the residual ion in an excited state (often called shake-up) or doubly ionized \cite{SvenssonJES1988,BeckerPRL1989,LablanquiePRL2000}.

It may be argued that the high temporal resolution achieved in attosecond experiments prevents any spectral accuracy and thus may affect the interpretation of experimental results. This is especially true when different processes can be induced simultaneously and lead to photoelectrons with kinetic energies within the bandwidth of the excitation pulse. In fact, the natural trade-off between temporal and spectral resolution may be overcome, as beautifully shown in the visible spectrum using high-resolution frequency combs based upon phase-stable femtosecond pulse trains \cite{MarianScience2004}.

\begin{figure}[htbp]
  \input{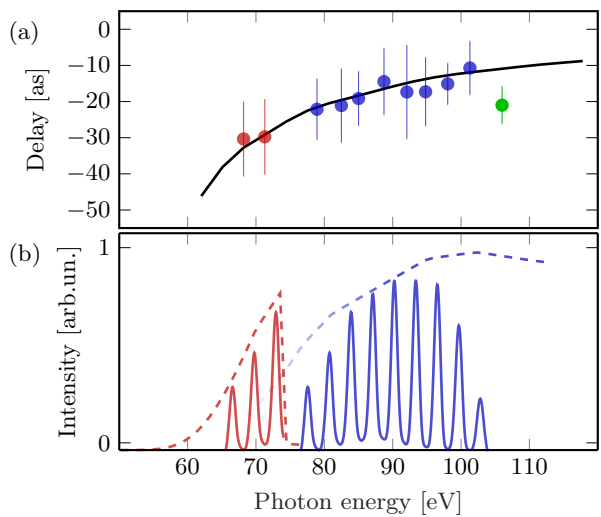}
  \caption{\textbf{Photoionization time delays in the $\mathbf{2s}$ and $\mathbf{2p}$ shells.} (a) Time delay differences [$\tau_\mathrm{A}(2\mathrm{s})-\tau_\mathrm{A}(2\mathrm{p})$] in neon as a function of photon energy for the two spectra shown in (b) (red and blue dots). Theoretical calculations within many-body perturbation theory (black solid line) agree very well with the experimental data. Also shown is the streaking experiment from \cite{SchultzeScience2010} (green dot).	(b) Photon spectra used in the measurement. High-order harmonics are generated in neon gas and filtered with a combination of 200~nm thick Al and Zr filters (red spectrum) and with two Zr filters (blue spectrum). The dashed lines illustrates the transmission curves of the two combinations of filters \cite{HenkeADNDT1993}.
 }
  \label{exp_delays}
\end{figure}

In this work, we bridge the gap between high-resolution photoelectron spectroscopy and attosecond dynamics, making use of the high-order harmonic spectrum obtained by phase-stable interferences between attosecond pulses in a train. We present a study of photoionization time delays of the $2\mathrm{s}$ and $2\mathrm{p}$ shells in neon over a broad energy range from 65 to 100 eV, using an interferometric technique combining high temporal (20~as) and spectral (200~meV) accuracy, originally introduced for characterizing attosecond pulses in a train \cite{PaulScience2001,MairesseScience2003} and called RABITT (Reconstruction of Attosecond Beating by Interference of Two-photon Transitions).
Remarkably, our temporal and spectral resolution depends only partly on the properties of the extreme ultraviolet (XUV) pulses. In the limit of long infrared (IR) pulses leading to trains with reproducible attosecond pulses, the temporal resolution is only limited by the stability of our interferometer and the resolving power of the electron spectrometer. In the present work, our spectral resolution, limited both by the harmonic bandwidths and by the spectrometer resolution, estimated to be $\simeq 200$ meV, allows us to disentangle direct $2\mathrm{s}$ ionization from shake up processes, where a $2\mathrm{p}$ electron is ionized while a second is excited to a $3\mathrm{p}$ state. As shown in Fig.~\ref{exp_delays}(a), our experimental results for the difference between $2\mathrm{s}$ and $2\mathrm{p}$ time delays, as indicated by the red and blue dots, agree very well with theoretical calculations performed within the framework of many-body perturbation theory (the solid black line). Our experimental observation of a shake up process due to electron correlation also provides a possible explanation for the discrepancy between the pioneering result of Schultze et al.~\cite{SchultzeScience2010} (green dot) and theoretical calculations \cite{MoorePRA2011,NagelePRA2012,DahlstromPRA2012}.

\textbf{Photoionization time delays}. In general, experimentally measured delays can be considered as the sum of two contributions, $\tau_{\textsc{XUV}} + \tau_\mathrm{A}$, where the first term is the group delay of the broadband excitation $\textsc{XUV}$ field \cite{MairesseScience2003} and the second term reflects the influence of the atomic system. To eliminate the influence of the excitation pulse, two measurements can be performed simultaneously, for example, on different ionization processes \cite{SchultzeScience2010,KlunderPRL2011,ManssonNatPhys2014} or in different target species \cite{PalatchiJPB2014,GuenotJPB2014}. This enables the determination of relative photoionization time delays.  Absolute photoionization delays can be deduced if we assume that one of the delays can by sufficiently accurately calculated to serve as an absolute reference  \cite{OssianderNatPhys2016}.

In nonresonant conditions, the atomic delay $\tau_\mathrm{A}$ can in turn be approximated as the sum of two contributing delays, $\tau_\mathrm{W} + \tau_{\mathrm{cc}}$. The first term is the group delay of the electronic wavepacket created by absorption of XUV radiation, also called photoionization time delay or, shortly, Wigner delay. Already in 1955, E. Wigner interpreted the derivative of the scattering phase as the group delay of the outgoing electronic wavepacket in a collision process  \cite{WignerPR1955}. This interpretation also applies to photoionization with a dominant outgoing channel, with a factor one half to account for the fact that photoionization is a half collision. The second term, $\tau_{\mathrm{cc}}$, is a correction to the photoionization time delay due to the interaction of the IR field with the Coulomb potential, which is required for the measurement. At high kinetic energies, larger than $\approx 10$ eV, $\tau_{\mathrm{cc}}$ can be accurately calculated using either the asymptotic form of the wave function \cite{DahlstromPCCP2012} or by classical trajectories \cite{NageleJPB2011}. The index ``cc'' refers to the fact that the involved IR transitions are between two continuum states. In other works \cite{NageleJPB2011, OssianderNatPhys2016}, it is denoted $\tau_{\mathrm{CLC}}$ (where the index is an abbreviation of ``Coulomb-laser coupling''). In the case of multiple angular channels  with comparable amplitude \cite{GuenotJPB2014}, as is the case close to resonances \cite{LucchiniPRL2015} or with angle-resolved detection away from the XUV light polarization axis \cite{HeuserPRA2016}, the separation of the two contributions $\tau_\mathrm{W}$ and $\tau_{\mathrm{cc}}$ may become ambiguous.
\begin{figure}[htbp]
  \input{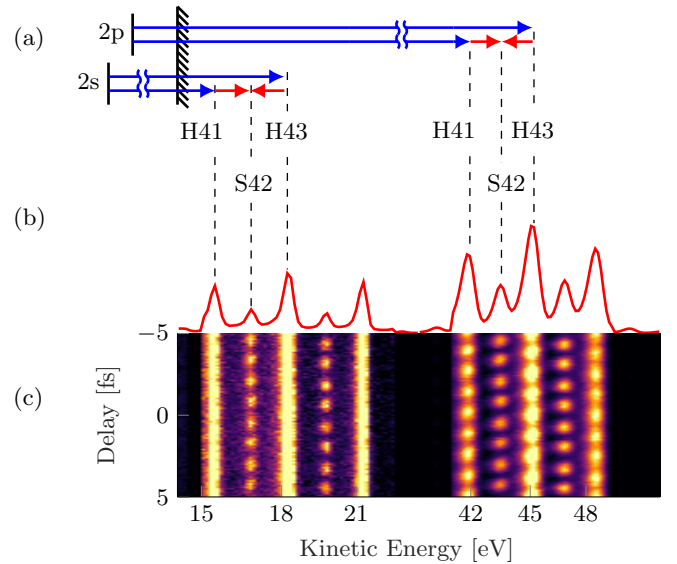}
  \caption{\textbf{Principle of the interferometric technique.} (a) Kinetic energy diagram for ionization from the $2\mathrm{s}$ and $2\mathrm{p}$ subshells using XUV (blue arrows) and IR (red arrows) radiation; (b) Time-averaged photoelectron spectrum obtained with Al-Zr-filtered harmonics. For both the $2\mathrm{s}$ and $2\mathrm{p}$ shell ionization results in three peaks due to absorption of harmonics (H41, H43 and H45) and two sidebands peaks (S42 and S44) reachable via two-color two-photon transitions. (c) Photoelectron spectrum as function of delay between the XUV pulse train and the IR field. The sideband amplitudes strongly oscillate as a function of delay. The electron yield from $2\mathrm{s}$ ionization has been multiplied by a factor of 5 for visibility.}
  \label{rabbitt}
\end{figure}

\begin{figure}[tbp]
  \input{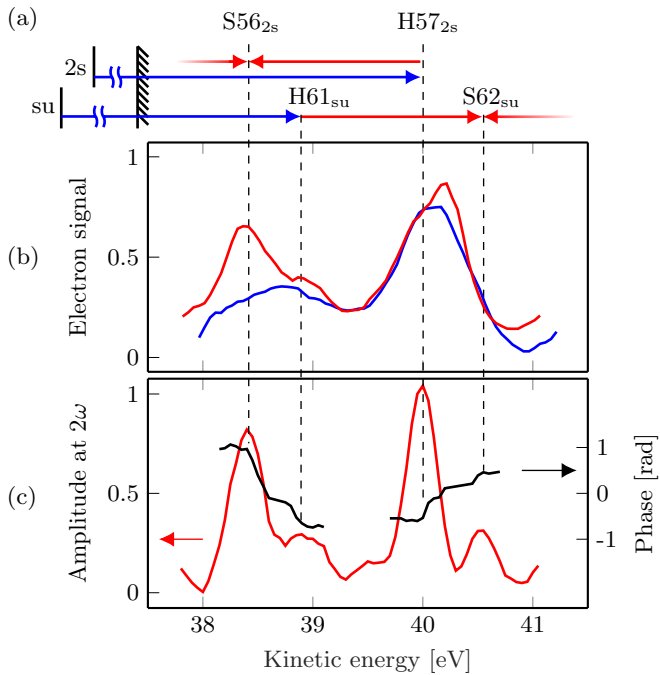}
  \caption{\textbf{Energy-resolved interferometric technique and identification of shake up process.} (a) Kinetic energy diagram for $2\mathrm{s}$ ionization and $2\mathrm{p}$-ionization accompanied by $2\mathrm{p} \rightarrow 3\mathrm{p}$ excitation (shake up).  The difference in threshold energy for these two processes is approximately 7.4 eV \cite{SvenssonJES1988}.
  (b) Photoelectron spectra for XUV only (blue) and XUV + IR (red). The electron peak due to shake-up induced by absorption of H61 partly overlaps with S56 from $2\mathrm{s}$-ionization. The shoulder on the S56 (red spectrum) can be attributed to one-photon induced shake up.
  (c) Energy-resolved amplitude and phase of the oscillation from the RABBITT-spectrogram. The harmonics oscillate out of phase with the sidebands, causing a sudden drop in the energy-resolved phase. The sideband originating from the shake-up state can be distinguished on the right side, allowing for a separate analysis of its time delay.}
  \label{shake_up}
\end{figure}

\textbf{Experimental method.}
The experiments were carried out using a Ti:Sapphire femtosecond laser system, delivering 20-fs pulses at 1 kHz repetition rate, 800 nm central wavelength with a pulse energy up to 5 mJ. The pulses are fed to an actively stabilized Mach-Zehnder interferometer  \cite{KroonOL2014}. In one arm, high harmonics of the fundamental laser frequency were generated from a pulsed gas cell filled with neon. The other arm contained a piezoelectric delay stage as well as a half-wave-plate and a broadband polarizer used for adjusting the probe pulse energy. Metallic filters placed in the XUV beam path limited the bandwidth of the XUV-pulses and eliminated the residual IR field present in the pump arm. Two sets of filters were used: a combination of zirconium and aluminum foils of 200 nm thickness each, yielding a narrow band-pass filter over the 60-75 eV range [red spectrum in Fig.~\ref{exp_delays}(b)] and a set of two Zr foils, resulting in a sharp edged high-pass filter above 70 eV [blue spectrum in Fig.~\ref{exp_delays}(b)].
The recombined pump and probe pulses were focused by a toroidal mirror into a magnetic bottle electron spectrometer similar to that described previously in \cite{ElandPRL2003}, with a 2~m long time-of-flight tube and a \mbox{$4 \pi$ sr} collection angle, and incorporating a set of retarding lenses. This spectrometer design combines a high collection efficiency with good spectral resolution ($\leq$ 100 meV) for low photon energies.

\textbf{Interferometric technique.}
Fig.~\ref{rabbitt} illustrates the principle of our interferometric measurement when using the Al-Zr filter combination. Two-photon ionization leads to sidebands which can be reached by two pathways: absorption of one harmonic and an IR photon, and by absorption of the next harmonic together with emission of one IR photon [Fig.~\ref{rabbitt}(a)]. Ionization of one sub-shell by the high-order harmonics and the IR field results in five electron peaks: three peaks due to single-photon ionization by harmonics 41, 43 and 45 and two sidebands 42 and 44. Since for this filter set the XUV spectrum spans less than $15$ eV and the difference in the ionization energies of the Ne $2\mathrm{s}$ and $2\mathrm{p}$ subshells is 26.8 eV \cite{SvenssonJES1988,BeckerPRL1989}, the spectra generated from the two subshells are energetically well separated [Fig.~\ref{rabbitt}(b)]. Fig.~\ref{rabbitt}(c) shows the variation of the spectrum as a function of the delay $\tau$ between the XUV and IR fields. The intensity of the sidebands oscillates according to \cite{KlunderPRL2011}
\begin{equation}
\label{SB}
S(\tau)=\alpha + \beta \cos[2\omega (\tau-\tau_\textsc{XUV}-\tau_\mathrm{A})],
\end{equation}
where $\alpha$ and $\beta$ are delay-independent and \mbox{$\omega$} denotes the IR frequency ($\pi/\omega=1.3$ fs in our experiment). Our analysis consists in determining the phase and amplitude of the signal oscillating at $2\omega$ by fitting Eq.~\ref{SB} to the experimental data. The delay $\tau_\textsc{XUV}$ depends only on the excitation pulse, which is the same for the $2\mathrm{s}$ and $2\mathrm{p}$-ionization paths. The difference in the photoionization time delays can therefore be obtained by comparing the oscillations of the sidebands corresponding to the same absorbed energy (e.g. S42), involving the same harmonics (H41 and H43). This analysis is performed over the bandwidth of the excitation pulse, from 60 to 75 eV in the experiment with the Al-Zr filters [red spectrum in Fig.~\ref{exp_delays}(b)] and from 80 to 100 eV using the Zr-filters [blue spectrum].

\textbf{Shake up.}
If the different energy components of the sideband are in phase, the analysis can be performed on the energy-integrated signal. In the present work, following the method described in \cite{GrusonScience2016}, we analyze the sideband oscillations across its spectrum, in steps of 50 meV. Fig.~\ref{shake_up} illustrates how this method allows us to identify shake up processes and eliminate their influence on the $2\mathrm{s}$-time delay measurement. In Fig.~\ref*{shake_up}(a), we indicate two competing ionization pathways leading to overlapping electron spectra: $2\mathrm{s}$-ionization by absorption of H57 and emission of one IR photon (S56); $2\mathrm{p}$-ionization and excitation $2\mathrm{p} \rightarrow 3\mathrm{p}$ by absorption of H61; Similarly, $2\mathrm{s}$-ionization by absorption of H57 and two-photon shake up (H61+IR) overlap.
Although a number of shake up processes come into play at photon energies above 50 eV, shake up to the  $2\mathrm{p}^4(^1\mathrm{D})3\mathrm{p}(^2\mathrm{P}^0)$ state, with binding energy equal to 55.8 eV, is the most intense \cite{SvenssonJES1988,FeistPRA2014}, reaching one sixth of the amplitude of $2\mathrm{s}$-ionization, and is thus comparable to a $2\mathrm{s}$-sideband. A comparison between the photoelectron spectra with and without IR shown in Fig.~\ref*{shake_up}(b) shows the effect of shake up on the right side of the $2\mathrm{s}$-sideband.
 In Fig.~\ref{shake_up}(c), the amplitude and phase of the $2\omega$ oscillation is shown as a function of energy. The phase is strongly modified in the region of overlap between $\mathrm{H}61_{\mathrm{su}}$ and $\mathrm{S}56_{2\mathrm{s}}$. In general, harmonic and sideband oscillate out of phase, so that, with poor spectral resolution, even a weak shake up harmonic signal strongly influences the phase of a partially overlapping $2\mathrm{s}$-ionization sideband signal. The spectrally-resolved phase of the $2\mathrm{p}$-sidebands (not shown) is completely flat, owing to the fact that this region is void of resonances \cite{LucchiniPRL2015} or shake-up states \cite{FeistPRA2014}. The time delays indicated in Fig.~\ref{exp_delays}(b) have been obtained by selecting a flat spectral region for the  $2\mathrm{s}$-phase determination, avoiding shake-up processes. We could also estimate the difference in time delay between shake-up and $2\mathrm{p}$-ionization to $-70\pm25$ as, by analyzing the shake-up sidebands amplitude and phase [see $\mathrm{S}62_{\mathrm{su}}$ on the right side in Fig.~\ref{shake_up}(c)].

\textbf{Comparison of theory and experiment.}
The key results obtained in the present work are summarized in Fig.~\ref{exp_delays}(a). For the experimental results [red and blue dots, corresponding to the spectra shown in (b)], the indicated error bars correspond to the standard deviation from ten spectrograms, weighted with the quality of the fitted sideband oscillations. The difference in time delay is negative, which indicates that $2\mathrm{p}$-ionization is slightly delayed compared to $2\mathrm{s}$-ionization, and decreases as the excitation energy increases. Unfortunately, we could not determine delays at energies higher than 100~eV due to overlap between electrons created by $2\mathrm{s}$-ionization with 100 eV photon energy and those by $2\mathrm{p}$-ionization with 70 eV. The difference in ionization energy between the two subshells corresponds almost exactly to 17$\omega$, which hinders any spectral analysis.

Fig.~\ref{exp_delays}(a) also presents calculations using a many-body perturbation theory approach for the treatment of electron correlation effects \cite{DahlstromJPB2014, DahlstromPRA2012}. Here, we calculate $\tau_\mathrm{A}$ by using lowest-order perturbation theory for the radiation fields. The interaction with the XUV photon is assumed to initiate the photoionization process with many-body effects included to the level of the random phase approximation with exchange. The laser photon is then assumed to act perturbatively on the photoelectron to drive a transition to an uncorrelated final state.
The final state is computed by solving an approximate Schr\"odinger equation with a static spherical potential of the final ion. Special care is taken that the laser dipole interaction of these two continuum waves is computed to radial infinity.
Using this method with ab-initio Hartree-Fock energies, it has been predicted \cite{DahlstromJPB2014} that the atomic delay from the $2\mathrm{p}$ state in neon is rather insensitive to interorbital correlation, while the coupling of the $2\mathrm{s}$ orbital is advanced by a few attoseconds due to coupling to the $2\mathrm{p}$ orbital. Here the calculations are improved further by using the experimental binding energies of $2\mathrm{p}$ and $2\mathrm{s}$.
The two-photon ionization amplitude is averaged over all emission angles ($\theta$) to mimic the experimental conditions. We emphasize that the excellent agreement obtained between theory and experiment for the difference in time delays between $2\mathrm{s}$ and $2\mathrm{p}$ ionization requires the careful energy-resolved analysis presented above and the disentanglement between $2\mathrm{s}$-ionization and shake up.

%FIGURE 4
\begin{figure}[tbp]
	\input{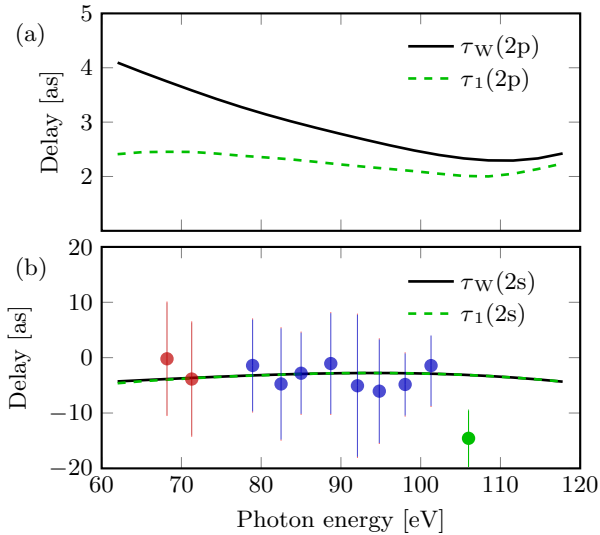}
	\caption{\textbf{Absolute photoionization time delays.}
		(a) Calculated $2\mathrm{p}$ Wigner delay $\tau_\mathrm{W}$ along the direction of the light polarization as a function of the photon energy (black solid line). The green dashed line indicates the angle-averaged one-photon ionization time delay accessible in the experiment. The difference between the two quantities is less than two attoseconds over the whole energy range. (b) Same as in (a) for $2\mathrm{s}$-ionization. The difference between $\tau_\mathrm{W}$ and $\tau_1$ is not visible. The experimental data (this work, red and blue dots and \cite{SchultzeScience2010}, green dot) is transformed to $\tau_1(2\mathrm{s})$ by subtraction of the analytical $\tau_{cc}$ and simulated $\tau_{1}(2\mathrm{p})$.}
	\label{theo_delays}
\end{figure}

\textbf{Absolute photoionization time delays.} Fig.~\ref{theo_delays} presents more details about the calculations and illustrates the contributions to the measured time delay differences.
In Fig.~\ref{theo_delays}(a), the black curve represents the Wigner delay $\tau_\mathrm{W}$ for $2\mathrm{p}$-ionization, calculated for an emission angle in the direction of polarization, while the dashed green curve is the angle-averaged time delay, defined as $\tau_1=\tau_\mathrm{A}-\tau_{\mathrm{cc}}$ (for the calculation of $\tau_{\mathrm{cc}}$, see \cite{DahlstromPCCP2012}).
The difference between the two curves is at most two attoseconds, which indicates a very small angle-dependence of the $2\mathrm{p}$ time delay \cite{LucchiniPRL2015}. Indeed, in this energy region, ionization towards the $\mathrm{s}$ continuum is much lower than towards the $\mathrm{d}$ continuum, which justifies our interpretation of $\tau_1$ in terms of Wigner delay for the $\mathrm{d}$-channel.
Fig.~\ref{theo_delays}(b) shows the same quantities for $2\mathrm{s}$-ionization.
Here, the difference between $\tau_1$ and $\tau_\mathrm{W}$ is not visible, which also justifies the interpretation of $\tau_1$ as Wigner delay. The red, blue and green dots have been obtained by subtracting from the experimental data [see Fig.~\ref{exp_delays}(a)] the calculated $\tau_\mathrm{A}(2\mathrm{p})=\tau_1(2\mathrm{p})+\tau_{\mathrm{cc}}(2\mathrm{p})$ and the continuum-continuum contribution $\tau_{\mathrm{cc}}(2\mathrm{s})$, thus extracting absolute Wigner delays for $2\mathrm{s}$-ionization.
The $2\mathrm{s}$ and $2\mathrm{p}$ ionization time delays at 100 eV are approximately -5 and +3 attoseconds, leading to a difference of -8 as. The energy-increasing, larger delays observed in Fig.~\ref{exp_delays}(a) reflect essentially the energy dependence of $\tau_{\mathrm{cc}}(2\mathrm{s})-\tau_{\mathrm{cc}}(2\mathrm{p})$, which itself is dominated by the variation of $\tau_{\mathrm{cc}}(2\mathrm{s})$.

Other calculations of the Wigner delays \cite{KheifetsPRA2013} agree to within a few as with our theoretical results and therefore with the experimental data. We have also compared our results with theoretical calculations in the conditions of a streaking experiment, i.e. with a stronger IR field and a single attosecond pulse~\cite{FeistPRA2014}. The calculated Wigner delay agrees very well with the data presented here.

In summary, we have presented experimental data and numerical calculations of the photoionization time delays from the $2\mathrm{s}$ and $2\mathrm{p}$ shells in neon for photon energies ranging from 65 eV up to 100 eV and retrieved the Wigner delay of the electronic $2\mathrm{s}$ wave-packet. The good agreement obtained gives us confidence in this type of measurement, and point out the necessity for keeping high frequency resolution in addition to high temporal resolution. We also carried out an energy-integrated instead of energy-resolved analysis of the sideband oscillations and obtained time delay differences which were often below those indicated in Fig.~\ref{exp_delays}(a), actually close to that retrieved by Schultze et al.~\cite{SchultzeScience2010}. This leads us to suggest that the discrepancy of the latter result with theory \cite{MoorePRA2011,NagelePRA2012,DahlstromPRA2012} might be due to the influence of shake up processes, not spectrally resolved in the experiment and not included in the theory (see, however, a detailed theoretical analysis including shake up processes in \cite{FeistPRA2014}). Our method can be significantly improved by using attosecond pulse trains generated with long laser pulses and/or in the mid infrared region. The long pulse duration allows the generation of stable attosecond pulse trains with many pulses (and thus narrow harmonic bandwidth) while the long wavelength leads to broad XUV spectra \cite{SheehyPRL1999,PopmintchevScience2012} and better energy sampling. The door is open to the study and control of photo-induced processes both in the time and frequency domain from the visible to the x-ray range.

%\lipsum[1]

\subsection*{Acknowledgements}
This research was supported by the European Research Council (Advanced grant PALP), the Swedish Research Council and the Knut and Alice Wallenberg Foundation. J.M.D. was funded by the Swedish Research Council, Grant No. 2014-3724.

%\newpage
%\bibliography{references,papers}

%\end{document}

\end{document}